\documentclass[aps,prl,reprint,superscriptaddress]{revtex4-1}

\usepackage{bbold}
\usepackage{graphicx}% Include figure files
\usepackage{dcolumn}% Align table columns on decimal point
\usepackage{bm}% bold math
\usepackage{epstopdf}
\usepackage{ulem}
\usepackage{color}
\usepackage{longtable} 
\usepackage{ulem}
\usepackage{amsmath}
\usepackage{amssymb}
\usepackage{sistyle}
\usepackage{bm}
\newcommand{\ket}[1]{$|#1\rangle$}

\newcommand{\er}{Er$^{3+}$ }
\newcommand{\ground}{$^4$I$_{15/2}$ }
\newcommand{\excited}{$^4$I$_{13/2}$ }
\newcommand{\YSO}{Y$_2$SiO$_5$ }

\newcommand{\rate}{s$^{-1}$}
\begin{document}

\title{Electron Spin Coherence in Optically Excited States of Rare-Earth Ions \\ for Microwave to Optical Quantum Transducers}

\author{Sacha Welinski}
\affiliation{PSL Research University, Chimie ParisTech, CNRS, Institut de Recherche de Chimie Paris, 75005, Paris, France}

\author{Philip J. T. Woodburn}
\affiliation{Department of Physics, Montana State University, Bozeman, MT 59717, USA}

\author{Nikolai Lauk}
\affiliation{Institute for Quantum Science and Technology and Department of Physics and Astronomy, University of Calgary, Calgary AB T2N 1N4, Canada}

\author{Rufus L. Cone}
\affiliation{Department of Physics, Montana State University, Bozeman, MT 59717, USA}

\author{Christoph Simon}
\affiliation{Institute for Quantum Science and Technology and Department of Physics and Astronomy, University of Calgary, Calgary AB T2N 1N4, Canada}

\author{Philippe Goldner}
\affiliation{PSL Research University, Chimie ParisTech, CNRS, Institut de Recherche de Chimie Paris, 75005, Paris, France}

\author{Charles W. Thiel}
\affiliation{Department of Physics, Montana State University, Bozeman, MT 59717, USA}

\date{\today}

\begin{abstract}

Efficient and reversible optical to microwave transducers are required for entanglement transfer between superconducting qubits and light in quantum networks. Rare-earth-doped crystals with narrow optical and spin transitions are a promising system for enabling these devices. Current approaches use ground-state electron spin transitions that have coherence lifetimes ($T_2$) often limited by spin flip-flop processes and spectral diffusion, even at very low temperatures. Here, we investigate spin coherence in an optically excited state of an  Er$^{3+}$:Y$_2$SiO$_5$ crystal at temperatures from 1.6 to 3.5 K for a low 8.7 mT magnetic field compatible with superconducting resonators. Spin coherence and population lifetimes of up to 1.6 $\mu$s and 1.2 ms, respectively, are measured by 2- and 3-pulse optically-detected spin echo experiments. Analysis of the decoherence processes suggest that ms $T_2$ can be reached at lower temperatures for the excited-state spins, whereas ground-state spin coherence lifetimes would be limited to a few $\mu$s for the same conditions due to resonant interactions with the other Er$^{3+}$ spins in the lattice and greater instantaneous spectral diffusion from RF control pulses. We propose a quantum transducer scheme with the potential for close to unit efficiency that exploits the advantages offered by spin states of the optically excited electronic energy levels.

\end{abstract}



%\keywords{}

\maketitle

Rare-earth (RE) doped crystals can exhibit long optical and spin coherence lifetimes at liquid helium temperatures that are promising for optical quantum technologies \cite{Thiel2011,Goldner2015,Zhong2015a}. For example, long lived quantum memories for light \cite{Laplane2017}, entanglement storage \cite{Clausen2011}, light to matter teleportation \cite{Bussieres2014}, as well as atomic gas to crystal quantum state transfer \cite{Maring2017} have all been demonstrated. Developments towards integrated and hybrid systems have also been reported using nano-structured materials \cite{Bartholomew2017,Zhong2017}, making RE-doped crystals a promising platform for solid-state quantum light-to-matter interfaces. In this respect, RE ions with an odd number of electrons are of particular interest since their strong magnetism offers the capability to efficiently couple to microwave photons. This could enable quantum-state transfer between superconducting qubits and light to build networks of quantum processors, for example \cite{Williamson2014,OBrien2014}. Strong coupling between superconducting resonators and RE ions has been demonstrated, and microwave to optical conversion efficiencies have been investigated \cite{Fernandez-Gonzalvo2015,Chen2016,Probst2015,Probst2013}.
% Long time quantum storage could also be possible by using isotopes possessing nuclear spins.  

Using both optical and spin transitions requires strong coupling between light and microwaves. Ensembles of RE ions in crystals are generally employed in applications to enhance the interaction efficiency with photons. While providing a number of advantages, ensembles also lead to specific difficulties when considering potential optical to spin coherence transfer processes. In particular, the  weak optical oscillator strengths limit addressable bandwidths for $\pi$ pulses to much less than 100 MHz; however, optical inhomogeneous linewidths are usually of the order of 1 GHz or larger \cite{Thiel2011}, which means that a majority of ions, and therefore spins, do not participate in the coherent excitation process and are only 'spectators'. When coherence is stored in a superposition of the ground-state electron spin states, two undesirable effects may appear. First, the excitation can be incoherently transferred by fast spin flip-flops, or spin diffusion, to other spins in the lattice, including the spectator spins, resulting in decoherence and loss of the stored quantum state. Secondly, to efficiently drive the spin transition, pulse bandwidths as large as the spin inhomogeneous linewidths are required since there is no correlation between optical and spin frequencies within their respective inhomogeneous linewidths. However, these high-bandwidth control pulses will also drive a large number of spectator spins, potentially causing decoherence by instantaneous spectral diffusion (ISD) from long-range spin-spin interactions \cite{Herzog1956,Thiel2014a,Probst2015}. Paramagnetic RE ions exhibit significant spin-spin interactions even for RE concentrations as low as 10 ppm \cite{Cruzeiro2017a}.  As a result, this source of decoherence often cannot be eliminated by reducing the ion concentration without reducing the optical absorption and limiting ultimate device efficiency. 

We propose to circumvent these limitations by using spin transitions for optically excited electronic states. Because the excited ions' spins are not resonant with the much larger number of spectator spins in the lattice, decoherence due to the flip-flop process will be significantly reduced.  Furthermore, only optically excited spins are resonant with the microwave fields, avoiding excitation of spectator spins by the control pulses and reducing ISD to the lowest possible extent. While this scheme is fundamentally limited by the excited-state population lifetime $T_{1,\mathrm{opt}}$, this can be longer than ms for RE ions, meeting the requirements for microwave-optical transduction protocols. A particularly favorable ion is erbium with typical excited-state lifetimes of $\sim$10 ms for the \excited level \cite{Bottger2006a}. This 1.5 
$\mu$m telecom wavelength transition has also attracted strong interest for building fiber-based quantum networks and transducers \cite{Saglamyurek2015,Chen2016,Probst2013}. Here, we investigate electron spin coherence in the excited state of an Er$^{3+}$ doped Y$_2$SiO$_5$ crystal, and observe, for the first time to our knowledge, excited-state electron spin echoes for RE ions.  We also present an excited-state transduction protocol that can reach close to unity efficiency. 

\begin{figure}
\includegraphics[width=\columnwidth]{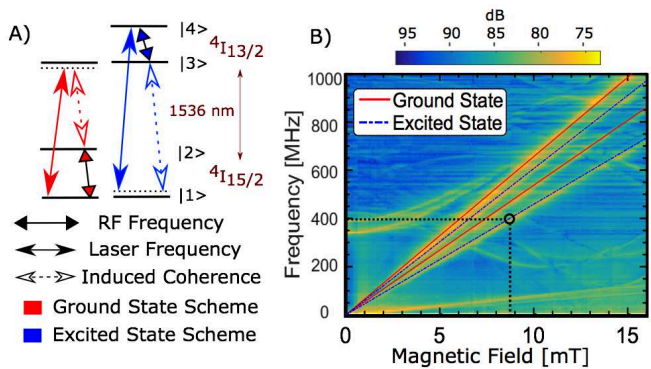}
\caption{A) Energy levels of \er  and RHS schemes for ground- and excited-state spin coherence studies (\ket{1}$\leftrightarrow$\ket{2} and \ket{3}$\leftrightarrow$\ket{4} are electron spin transitions). B) RHS measurements at 3 K on \er at site 1 of \YSO. Lines: fitted Zeeman transition frequencies for the zero-nuclear-spin Er$^{3+}$ isotopes. Circle: spin echo experimental condition.}
\label{raman_heterodyne}
\end{figure}

Experiments were performed using coherent Raman heterodyne scattering (RHS) techniques \cite{Mlynek1983} for the three-level schemes shown in Fig. \ref{raman_heterodyne}. This method employs a radio-frequency (RF) field resonant with a Zeeman transition \ket{1}$\leftrightarrow$\ket{2} or  \ket{3}$\leftrightarrow$\ket{4} to create a ground- or excited-state electron spin coherence (Fig. \ref{raman_heterodyne}A). 
A resonant laser beam drives the optical transition \ket{1}$\leftrightarrow$\ket{3} or \ket{1}$\leftrightarrow$\ket{4}. The combined optical and RF coherences then induce a coherence on the other optical transition \ket{2}$\leftrightarrow$\ket{3} or \ket{1}$\leftrightarrow$\ket{3} that then interferes with the laser field, producing a beat note at the \ket{1}$\leftrightarrow$\ket{2} or  \ket{3}$\leftrightarrow$\ket{4} frequencies that can be detected using a fast photodiode. 

We use a 50 ppm \er-doped YSO sample grown by Scientific Materials Corp. and cut along the three dielectric axes: $b$, $D_1$, $D_2$ \citep{Li1992}. \er can substitute for Y$^{3+}$ at two inequivalent crystallographic sites (denoted 1 and 2). Furthermore, each site has two sub-groups with different local orientations that exhibit different Zeeman effects unless the magnetic field is applied either parallel or perpendicular to the $b$ axis. The sample was mounted in an Oxford Optistat helium cryostat with a magnetic field applied along $D_1$ using a Helmholtz coil.  
An external-cavity diode laser was set at 1536.49 nm (vacuum) in resonance with the transition between the lowest crystal field levels of the $^4$I$_{15/2}$ and $^4$I$_{13/2}$ multiplets for ions at site 1 \cite{Bottger2006a}. The laser was amplified and then focused into the crystal with propagation along the crystal's $b$ axis and polarization along the $D_2$ axis. The CW laser incident on the crystal produced nearly equal populations in both the ground and excited states over $\sim$1 MHz bandwidth. Optical pumping also induced population differences within the ground- and excited-state Zeeman sub-levels necessary to detect spin coherence with the RHS method. Transmitted light was detected by an AC-coupled photoreceiver with 1 GHz bandwidth. RF pulses with magnetic field amplitudes of up to several Gauss were applied along $b$ through a copper wire RF waveguide held next to the crystal surface. 

% A spectrum analyzer (see supplementary) generates the RF signal (-10 dBm output) at $\omega_H$ varying from 0 to 1 GHz.  This field is amplified (+33 dB) before being sent into the cryostat. The sample, a 3.9 mm-long along b axis \YSO single crystal doped with 50 ppm of \er, sits on a copper wire through which the RF power circulates, so that the RF excitation is along the b axis of the crystal (see supplementary). The RF wire is terminated by a 50 $\Omega$ impedance external to the cryostat in order to avoid any back-reflection. The laser beam was continuously sent through the sample ($\mathbf{k}$ along b axis) during the measurements. The optical detection after the sample was made by a (New Focus model 1611) 1 GHz bandwidth low noise photodiode and the acquired signal was sent back into the spectrum analyzer in order to down-mix the signal and to measure the Raman beat amplitude. 
%After each RF frequency sweep, the magnetic field was ramped up in order to get the Figure \ref{raman_heterodyne}B). 

Fig. \ref{raman_heterodyne}B shows the RHS spectra for frequencies of up to 1 GHz as a function of magnetic field strength. In these experiments, an RF spectrum analyzer was used to generate the constant RF excitation and to analyze the photoreceiver signal (see Supplemental Material \cite{SM}). 
%Transitions could be observed with signal to noise ratios (SNR) up to 20 dB, enabling determining resonance frequencies with a precision of XX. With the available magnetic field strength, up to 25 mT, we could to measure transitions up to 1.8 GHz but with a lower SNR due to the limited bandwidth of our detector. This technique is therefore useful to investigate a broad frequency range, including that of superconducting resonators used for single microwave photons. We could also record excited state spin transitions thanks to the continuous optical field thad averaged to about t created a large population in the excited state. 
The four straight lines observed in Fig. \ref{raman_heterodyne}B correspond to electron-spin transitions for the ground and excited states of \er isotopes with zero nuclear spin, as deduced from the magnetic field direction and known \textbf{g} tensors \cite{Guillot-Noel2006,Sun2008}. Two transitions are observed for each state since the 
magnetic field was not exactly parallel to $D_1$, resulting in two inequivalent sub-groups for each site. 
The corresponding effective ground-state values $g_{g}$ are 4.75 and 3.85 ($\pm$ 0.3), and the excited-state values $g_{e}$ are 4.35 and 3.27  ($\pm$ 0.3), depicted in Fig. \ref{raman_heterodyne}B). 
% These measured values allow us to determine the exact direction of the magnetic field with respect to the $bD_1D_2$ crystal frame: 
% $\theta=86.91 \pm 0.01\arcdeg $ and $\phi=7.0\pm 0.2 \arcdeg$. 
All spin transitions exhibited linewidths of $\sim$10 MHz, similar to those previously reported \cite{Probst2013,Guillot-Noel2006}. No significant variation in linewidth with magnetic field strength was observed, indicating that broadening of the spin transitions does not arise from inhomogeneity in the \textbf{g} tensors. 
Other transitions with more complex field dependence are also visible in Fig. \ref{raman_heterodyne}B) and are attributed to hyperfine transitions of the $^{167}$\er isotope ($I=7/2$, abundance 22.9 \%).

\begin{figure}
\includegraphics[width=1\columnwidth]{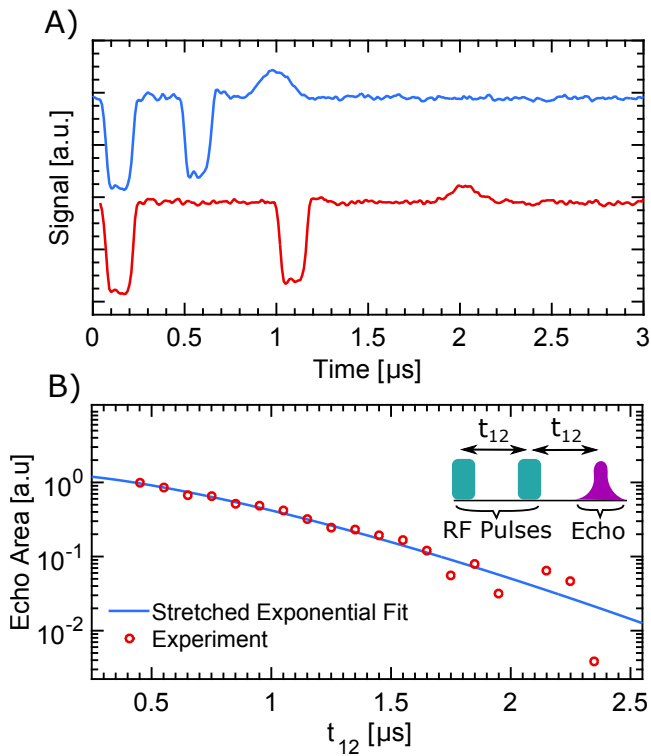}
\caption{A) Examples of optically detected electron spin echoes in the \excited excited state for different pulse delays with $T= 1.9$ K and $B =  8.7$ mT.  B) Measurement (circles) and fit (line) of echo area decay at 1.9 K, giving $T_2=1.6 \pm 0.2$ $\mu$s and $x=1.4 \pm 0.2$. Inset: RF pulse sequence.}
\label{echo}
\end{figure}

Excited-state spin echoes were measured using RF pulses generated by a gated source. The photoreceiver signal was amplified, filtered, and then down mixed with a local oscillator (see Supplemental Material \cite{SM}). The DC magnetic field of 8.7 mT was applied along the same orientation as in the CW experiments, and an RF frequency of 400 MHz was used to study the excited-state transition (Fig. \ref{raman_heterodyne}C) for temperatures from 1.6 to 3.5 K.

The signal for the 2-pulse echo sequence is shown in Fig. \ref{echo}A for two different delays between the excitation pulses. Pulse lengths of 150 ns were used to maximize the echo signal. The detected echo has a $\pi$ phase shift relative to the excitation pulses, confirming that the entire sequence was phase coherent. The large excited-state population created by CW laser excitation allowed strong echo signals to be produced. 
By varying the delay between the pulses, we measured the decay of the integrated echo signal area (Fig. \ref{echo}B). The decay was fitted with a Mims shape \cite{Mims1968}:
$
A(\tau)=A\exp{[-(2t_{12}/T_{2e})^x]}
\label{streched}
$
where $T_{2e}$ is the 1/e phase coherence lifetime and $t_{12}$ is the delay between excitation pulses. The extracted $T_{2e}$ at 1.9 K was 1.6 $\pm\ 0.2$ $\mu$s (200 kHz homogeneous linewidth), with $x = 1.4 \pm  0.2$.
The non-exponential behavior indicates a spectral diffusion effect due to interactions with the bath of \er spins in the lattice \cite{Mims1968}. The effect of $^{89}$Y nuclear spins is expected to be much smaller because of their weak magnetic moment and slow flip rates \cite{Bottger2006b}. Due to the narrow optical excitation bandwidth, the excited-state ion concentration for site 1 is  $\sim$10$^{3}$ times lower than the ground-state concentration so that they do not contribute significantly to spectral diffusion. At the \er concentration and low fields studied here, ground-state spins are only weakly polarized for both sites and are expected to relax mainly by mutual spin flip-flop processes (spin diffusion), even at very low temperatures.

%More precisely, the flip rate $R$ of the spins in this bath should obey $Rt_{12}\ll 1$ for the delays used in our experiments. 

%We divide Er spins into three groups. A are the probed spins, i.e. excited state spins in site 1; B are site 1 ground state spins and C are ground state site 2 spins. 
%We were not able to observe any echo in the optical ground state at a frequency of XX MHz. 
%
% This may be due to a too short spin memory time due to a short spin lifetime time. Indeed, we expect the optical ground state to be more perturbed by resonant spin flip-flop with the remaining spins in thermal equilibrium. As a proof of a short spin lifetime time in the optical ground state, we did not manage to burn any a hole in the optical absorption line at those conditions. \textcolor{red}{But why is T$_{Zeeman}$ so short?}.
% At the opposite, both spin states of \excited are empty at thermal equilibrium. It is though easy to get a decent population difference when we excite selectively one of the four optical transitions (\ket{1}$\leftrightarrow$\ket{3}, \ket{1}$\leftrightarrow$\ket{4}, \ket{2}$\leftrightarrow$\ket{3} and \ket{2}$\leftrightarrow$\ket{4}).\\

%This variation is characteristic of significant spectral diffusion (cf Mims) of the spins between the excitation pulses. 

\begin{figure}
\includegraphics[width=1\columnwidth]{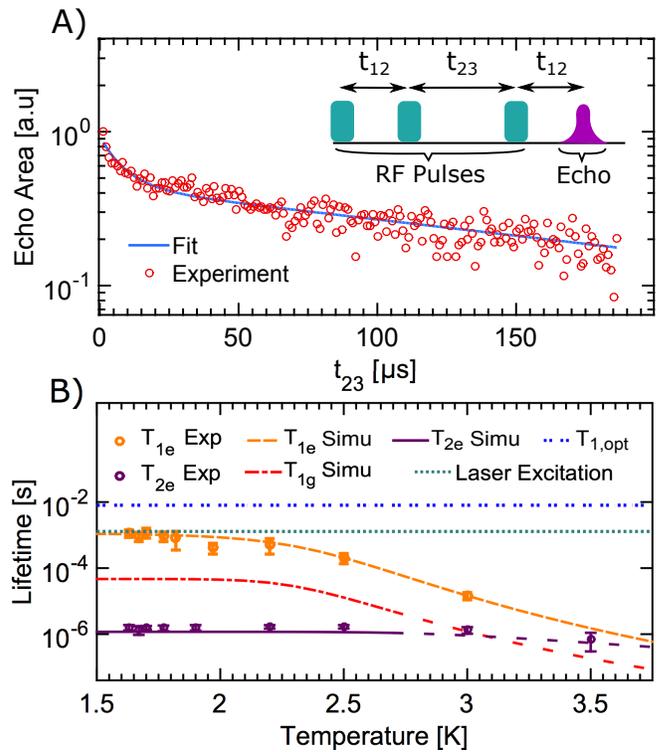}
\caption{A) Excited-state stimulated spin echo decay measured at 2.5 K (circles) and fit to the spectral diffusion model (line, see text).  Inset: RF pulse sequence. B) Experimental and modeled coherence and population lifetimes as a function of temperature.}
\label{temperature}
\end{figure}

To probe the decoherence mechanisms, we measured 3-pulse stimulated spin echoes using the pulse sequence shown in Fig. \ref{temperature}A. Stimulated echo measurements allow both spin relaxation and spectral diffusion dynamics to be studied over the timescale of $T_1$, whereas 2-pulse echoes are limited to the much shorter $T_2$ time scale \cite{Mims1968,Bottger2006b}.
We measured 3-pulse echo decays as a function of $t_{23}$, with $t_{12}$ fixed at 0.3  $\mu$s for temperatures between 1.6 and 3 K. An example decay is shown in Fig. \ref{temperature}A. The initial fast decay is due to spectral diffusion while the slower exponential decay component results from population relaxation.

Effects of spectral diffusion on echo decays can be modeled by a time-dependent effective homogeneous linewidth $\Gamma_{\mathrm{eff}}$ \citep{Bottger2006b}, with the echo amplitude given by
\begin{equation}
A(t_{12},t_{23}) = A_0 e^{(-\frac{t_{23}}{T_{1e}})}e^{[-2t_{12}\pi\Gamma_{\mathrm{eff}}(t_{12},t_{23})]}.
\label{tpecho}
\end{equation}
The excited-state spins probed in the echo sequence are perturbed by the bath of ground-state Er$^{3+}$ spins that relax at a rate $R$ with a distribution of interaction strengths characterized by  $\Gamma_{\mathrm{SD}}$ so that the effective linewidth in Eq. \ref{tpecho} can be written as \cite{Bottger2006b}
\begin{equation}
\Gamma_{\mathrm{eff}}(t_{12},t_{23}) = \Gamma_0+\frac{1}{2}\Gamma_{\mathrm{SD}}(Rt_{12}+1-e^{-Rt_{23}}).
\label{sd}
\end{equation}
To reduce the number of fit parameters, $T_{1e}$ was determined from an exponential fit to the tail of the 3-pulse echo decays. Moreover, at the temperatures and magnetic field used here, $\Gamma_{\mathrm{SD}}$ and the flip-flop rate $R_{\mathrm{ff}}$ are temperature independent so that the relaxation rate $R$ can be modeled by
$
R = R_{\mathrm{ff}}+\alpha_{O,g} / [ \exp \left( \Delta_g/ T \right)-1]
$
as a function of temperature $T$. The second term corresponds to the resonant two-phonon Orbach process where $\Delta_g$ is the energy of the next crystal field level above the ground state (in Kelvin) and $\alpha_{O,g}$ is the coupling strength. The two-phonon Raman and one-phonon direct terms are both negligible for our conditions. $\Gamma_0$ is also taken as independent of temperature. Fits of this model to the experimental  2- and 3- pulse echo decays (see Supplemental Material \cite{SM}) give the following values: $\Gamma_0 = 2.7 \times 10^5$ Hz, $\Gamma_{\mathrm{SD}} = 4.3 \times 10^5$ Hz, $R_{\mathrm{ff}} = 2.1 \times 10^4$ \rate, $\alpha_{O,g} = 50 \times 10^{10}$ Hz and $\Delta_g = 40$ K.

The experimental parameters compare well with theoretical estimates for decoherence from spin flip flops of site 1 \er ground-state spins (see Supplemental Material \cite{SM}), with calculated values of $\Gamma_{\mathrm{SD, theory}} = 4.4 \times 10^5$ Hz and  $R_{\mathrm{ff,theory}} = 1.6 \times 10^5$ \rate. The crystal field level splitting of 57 K \cite{Bottger2006a} is larger than the fitted value, a common observation in spin-lattice relaxation studies \cite{Young1966,Wolfowicz2015,Bottger2016}. For our field orientation, the ground-state $g$ factors for site 1 and 2 spins are 4 and 14 \cite{Sun2008} so that the site 2 flip-flop rate will be faster than for site 1 by a factor of roughly $(14/4)^4= 150$.  This much faster rate causes decoherence due to site 2 spins to be reduced because of the well-known "motional narrowing" effect \cite{Bloembergen48}. Consequently, we attribute $\Gamma_0$ to site 2 ground-state spins that produce decoherence over sub-$\mu$s timescales. This conclusion is supported by a simple theoretical estimate of $\Gamma_{\mathrm{0,theory}} =6 \times 10^5$ Hz for this effect that is consistent with the observed value (see Supplemental Material \cite{SM}).

Fig. \ref{temperature} shows the calculated variation of $T_{1g}=1/R$ as a function of temperature. Below 2.2 K, relaxation is dominated by the flip-flop process. A plot of $T_{1e}$ extracted from Eq. 1 and the curve calculated from Eq. \ref{sd} by setting $t_{23}=0$ is shown in Fig. \ref{temperature}B. The $T_{1e}$ variations are explained by the sum of the optical excitation and emission rates combined with Raman and Orbach spin relaxation processes (see Fig. 3B and Supplemental Material \cite{SM}). At the lowest temperatures, $T_{1e}$ reached 1.2 ms, limited by the optical stimulated emission rate due to continuous excitation by the laser. In a pulsed configuration without laser excitation during the echo sequence, this would increase to the limit of $T_{\mathrm{1,opt}}=8$ ms \cite{thiel2012}.

For our conditions, $T_{2e}$ is limited by decoherence due to relaxation of ground-state spins. This decoherence would be reduced at lower temperatures since the Orbach and Raman contributions rapidly decrease as $\exp(-\Delta_e/T)$ and $T^9$, respectively, while flip-flop rates decrease as
$
[\mathrm{sech}( g_{\mathrm{eff}}\mu_B B/2kT )]^2
$,
where $k$ is the Boltzmann factor.  For example, consider a weak field of 50 mT that is compatible with superconducting resonators and where the excited-state splitting is about 2 GHz, also in the range of typical microwave photons. At a temperature of 20 mK, decoherence only results from site 1 ground-state spin flip-flops because site 2 spins, with their large $g$ factor, are completely polarized. The site 1 flip-flop rate $R_{\mathrm{ff}}$ would be reduced by a factor of $\sim$350 compared to 2 K.  Together, these effects would result in a much longer excited-state spin coherence lifetime of $T_{2e} \approx 1.8$ ms (see Supplemental Material \cite{SM}). This is likely to approach the decoherence limit due to $^{89}$Y spin flips \cite{Bottger2006b}. In contrast, even at very low temperatures, site 1 ground-state excitations will still experience rapid decoherence through the flip-flop process due to the large number of other resonant spins present in the lattice. In fact, spins excited into the higher energy spin state will have an increasing number of neighbors in the lower energy spin state that they can flip flop with as the temperature is decreased, accelerating decoherence to as much as twice the high-temperature rate. For our system, this effect will limit $T_{2g}$ to less than 30 $\mu$s even at the lowest temperatures. Moreover, a rephasing control pulse applied over the entire spin linewidth would cause strong ISD, reducing $T_{2g}$ to $\sim$1 $\mu$s, independent of temperature (see Supplemental Material \cite{SM}). These effects explain why we did not observe any ground-state spin echo for the conditions used in the excited-state spin echo measurements. We note that our analysis is also in qualitative agreement with the ground-state site 2 coherence lifetime of $5.6$ $\mu$s that has been observed at 30 mK with a different magnetic field orientation \cite{Probst2013}.

Finally, we turn to an excited-state scheme for an optical to microwave transducer.  Previous proposals used the ground-state spin by coupling it off-resonantly \cite{Williamson2014} or resonantly \cite{OBrien2014} to a microwave cavity. One limiting factor for efficient conversion in \cite{OBrien2014} is the short coherence lifetime compared to the coupling strength. 

To exploit the potential increase in coherence time for the excited state, the protocol from \cite{OBrien2014} can be modified in the following way.  The first step is the same as in \cite{OBrien2014}, we prepare a narrow spectral feature and then apply a magnetic field gradient to produce an inhomogeneously broadened feature. An incoming optical photon is then absorbed on the \ket{1} - \ket{4} transition. The induced inhomogeneous broadening and the free evolution of the system lead to a dephasing of the optical coherence, preventing re-emission and ensuring that the optical photon is stored as a matter excitation. In the next step, we apply a $\pi$-pulse on the \ket{1} - \ket{3} transition, bringing the population from state \ket{1} to \ket{3}. The subsequent free evolution will further de-phase the system due to the inhomogeneous broadening of the spin state, but now at a possibly different rate. After a delay, we apply a second $\pi$-pulse to bring population back into \ket{1}, while simultaneously reversing the field gradient to begin the rephasing procedure. Once the dephasing due to inhomogeneous broadening of the excited state is compensated, we apply another $\pi$-pulse, moving the population to state \ket{3} to complete the rephasing procedure, leaving the system in a collective state that strongly couples to a microwave cavity. Assuming the same spin-cavity coupling strength of $v/2\pi = 34\ \text{MHz}$ as in \cite{Probst2013}, which is justified since the principal values of the magnetic $g$ tensors for the \ground and \excited states in \er:\YSO are roughly the same \cite{Sun2008}, and using our measured spin coherence lifetime of $T_{2e}= 1.6\ \mu s$, we can estimate the conversion efficiency using Eq. 8 in \cite{OBrien2014} to be $\eta \gtrsim 99\%$.
A further advantage of using the excited-state spin transition is the fact that only optically excited ions, i.e. those in the laser beam cross section, will interact with the microwave cavity, making spatial hole burning, which might be required in the previous proposals, superfluous. We note that the proposed protocol can also be reversed, i.e. it allows conversion of a microwave photon to a propagating telecom photon. Moreover, the bandwidth of the optical photon can be tuned in the protocol by controlling the strength of the field gradient.

In conclusion, we observed electron spin echoes in the optically excited state of an erbium doped crystal. Coherence lifetimes of up to 1.6 $\mu$s were recorded for a field of 8.7 mT at 1.9 K, and a detailed analysis of the decoherence processes suggest that ms lifetimes could be reached for conditions used in superconducting qubit studies. We propose a scheme to exploit these long coherence lifetimes for reversible optical to microwave conversion, with our analysis predicting near unity conversion efficiency. Overall, the possibility of using excited-state spin transitions opens a new and attractive way to coherently interface RE ensembles with microwave cavities and may stimulate new proposals for transducer devices.

%\appendix
%\section{}


We thank J. Bartholomew, N. Sinclair, M. Falamarzi, D. Oblak, and W. Tittel for useful discussions, and A. Marsh and R. Nerem for assistance during measurements. This work received funding from the joint French-US ANR project DISCRYS (No. 14-CE26-0037-01) and US NSF grant no. CHE-1416454, as well as Nano'K project RECTUS, the University of Calgary, and NSERC.

{\it Note:} A related experiment using hyperfine states of $^{167}$\er has been performed in parallel \cite{Rakonjac2018}.

\bibliography{library}

\end{document}